\documentclass[twocolumn,prl,showpacs]{revtex4}

\usepackage{graphicx}
\usepackage{rotating}
\usepackage{amsmath}
\usepackage{amsfonts}
\usepackage{amssymb}
\usepackage{enumerate}
\usepackage{longtable}
\usepackage{dcolumn}
\usepackage{bbm}

\begin{document}

\title{Concatenated Control Sequences based on Optimized Dynamic Decoupling}

\author{G\"otz S. Uhrig \footnote{On leave from
Lehrstuhl f\"{u}r Theoretische Physik I,
Technische  Universit\"{a}t Dortmund,
 Otto-Hahn Stra\ss{}e 4, 44221 Dortmund, Germany}}
\email{goetz.uhrig@tu-dortmund.de}
\affiliation{School of Physics, University of New South Wales, 
Kensington 2052, Sydney NSW, Australia}

\date{\rm\today}

\begin{abstract}
Two recent developments in quantum control, concatenation
and optimization of pulse intervals, are combined 
to yield a strategy to suppress unwanted couplings in quantum systems
to high order. Longitudinal relaxation and transverse dephasing can be 
suppressed so that systems with a small splitting 
between their energy levels can be kept isolated from their environment. 
The required number of pulses grows exponentially
with the desired order but is only the square root 
of the number needed if only concatenation is used.
An approximate scheme even brings the number down
to polynomial growth.
The approach is expected to be useful for quantum information and for
high-precision nuclear magnetic resonance.
\end{abstract}



\maketitle

If one wants to measure small shifts in local energy levels high-precision
nuclear magnetic resonance (NMR) is a powerful means of investigation.
In order to reduce the line widths which limit the achievable 
resolution it is important to isolate the system under study as much as 
possible from its environment. It is common in NMR to do this by
intricate sequences of pulses, see e.g.\ Ref.\ \onlinecite{haebe76}.

Similarly, one of the big hurdles in the realization of quantum information 
(QI) is to keep information stored without loss due to decoherence. 
Besides quantum error correction, see e.g.\ 
\cite{stean98,zolle05}, dynamic decoupling  (DD) is one important tool to
reduce decoherence \cite{ban98,viola98,viola99a,facch05,khodj05,uhrig07}.
Here, as for NMR, a sequence of pulses is used to achieve an effective
Hamiltonian where the quantum bit (qubit) is as much isolated from its
environment as possible.

So in both these fascinating areas of physics the design of
appropriate pulse sequences is of great relevance.

The fundamental idea is to compensate the effect of 
a Hamiltonian (or a part of it) by inverting its sign by a $\pi$ pulse.
This is known since long
as spin-echo technique \cite{hahn50} which works very nicely
for static magnetic fields. Its sequence reads 
$p_\mathrm{SE}=f_\tau\,\pi\,f_{\tau}$.
 Further improvement is obtained by passing
to the combination of two pules in $p_\mathrm{CPMG}=
f_\tau\,\pi\,f_{2\tau}\,\pi\,f_\tau$
\cite{carr54,meibo58} which is commonly called CPMG after the first letters
of its inventors. The total duration of the cycle is $t =4\tau$.
Here $\pi$ stands for an ideal, instantaneous $\pi$ pulse which takes
a part of the Hamiltonian to its negative while $f_\tau$ stand for the time
evolution due to the unchanged Hamiltonian over the interval $\tau$.
Both basic cylces are commonly iterated: the iteration
of the spin echo leads to $p_\mathrm{PDD}=f_\tau\,(\pi\,f_\tau)^n$ with 
$t =(n+1)\tau$ which is often called periodic dynamic decoupling
(PDD). The iteration of the CPMG leads to 
$p_\mathrm{iCPMG}=(f_\tau\,\pi\,f_\tau)^{2n}$ with 
$t =2n\tau$ which is successful in many circumstances, see 
for instance Refs.\ \onlinecite{witze07a,yao07}. 
It is particularly robust against
soft high-energy cutoffs \cite{uhrig08,cywin08}.

Recently, two powerful extensions of DD have been proposed.
The first is concatenation \cite{khodj05,khodj07} (CDD). By iterating
the concatenation of pulse sequences $p_{n+1}=p_n\,X\,p_n\,Z\,p_n\,X\,p_n\,Z$
with $p_0=f_\tau$ Khodjasteh and Lidar could show that the
interaction between a qubit and its environment can be reduced
to order $t ^{n+1}$. Hereby $X$ stands for a $\pi$ pulse about
$\sigma_X$, the Pauli matrix of the qubit in $x$-direction, and
$Z$ stands for a $\pi$ pulse about $\sigma_Z$.
The caveat of this approach is that the number of necessary pulses
grows exponentially with a high power $\propto 4^n$ with the 
concatenation level $n$, thus with the power in $t $.
Hence the implementation of a CDD sequence for a given order can be
technically very demanding.

The second extension is the optimization of the intervals between
the pulses. They need  not be chosen equidistant and this additional
degree of freedom can be exploited for optimization \cite{uhrig07}
leading to the UDD sequence.
In a dephasing model where bosons are coupled linearly to the qubit
(spin-boson model) 
the author showed that the instants $\tilde\delta_j$
($j\in \{1,2,\ldots n\}$) for $\pi$ pulses are optimum if
\begin{equation}
\label{eq:udd0}
\tilde\delta_j = t  \sin^2(\pi j/(2(n+1))).
\end{equation}
Based on the observation that this formula does not depend on any details of
the system and on the knowledge that most degrees of freedom 
behave at high temperatures like gaussian fluctuations the author
concluded this sequence is not specific to the spin-boson model.
The main advantage of $p_\mathrm{UDD}$ as defined by \eqref{eq:udd0}
is that one gains an order in $t $
with each pulse added \cite{uhrig07,uhrig08}. So no exponential costs
arise. 

Up to $n=5$ the intervals of the UDD sequence were found even earlier
by explict general calculation, not relying on a specific model \cite{dhar06}.
Based on numerical evidence and on analytical evaluation of the first
orders up to $n=14$ by large-scale recursions
it was subsequently conjectured that
\eqref{eq:udd0} applies to all pure dephasing models \cite{lee08a,uhrig08}. 
It is not specific to the spin-boson model at all. Very recently, this
well-founded hypothesis could be turned into a theorem by mathematical
proof \cite{yang08}. The caveat of this approach so far is that
it can only deal with pure dephasing or longitudinal relaxation, respectively.
In particular for so-called low-field systems, where the splitting between
the two energy levels of the qubit is small, this is a serious drawback.
In such systems, the commonly used rotating-frame approximation works only
poorly and so dephasing and spin flips have to be taken into account.

So we are faced with two interesting sequences, CDD and UDD, both with
strong points (CDD can eliminate all couplings, UDD requires only linear
number of pulses) and weak points (CDD requires exponentially large
number of pulses, UDD eliminates only pure dephasing). In the present
work, we show that their combination leads to the partial
combination of their advantages.

In the following derivation we will consider ideal, instantaneous
$\pi$ pulses for simplicity. Such pulses can be approximated
to some extent  by very short pulses (but see also the 
results in Refs.\ \onlinecite{li07,li08a}) and, more efficiently, 
by suitably shaped pulses \cite{pasin08a,karba08}.

To derive the advantages of a concatenated UDD sequence (CUDD) we
first recall that a UDD sequence
of $n$ pulses with $\pi$ rotations ($Z$) about the $z$ direction
suppresses the relaxation along $z$ \cite{yang08}, 
i.e., very little spin flips occur.
Formally we start from the most general decoherence Hamiltonian
of a single qubit
\begin{equation}
H = \sum_{\gamma\in\{0,x,y,z\}} \sigma_\gamma\otimes A_\gamma,
\end{equation}
where $\sigma_0$ is the identity in the Hilbert space of the qubit
and the $A_\gamma$ are operators of the environment only. Then
the time evolution $p^n_\mathrm{UDD}$ of the  UDD sequence 
\begin{equation}
\label{eq:udd1}
p^n_\mathrm{UDD} = f_{t -\tilde\delta_{n}}Z\, 
 f_{\tilde\delta_{n}-\tilde\delta_{n-1}}Z\ldots
Z\, f_{\tilde\delta_{3}-\tilde\delta_{2}} 
Z\, f_{\tilde\delta_{2}-\tilde\delta_{1}} Z \, 
f_{\tilde\delta_{1}}
\end{equation}
behaves  as if an effective Hamiltonian 
$H^\mathrm{eff}$ acted 
\begin{equation}
\label{eq:efftimevol}
p^n_\mathrm{UDD} = \exp(- i t  H^\mathrm{eff})
+{\cal O}(\alpha t ^{n+1})
\end{equation}
where $\alpha:=\max_\gamma(||A_\gamma||)$ is used; any
 operator  norm $||\cdot||$ can be employed.
The effective Hamiltonian  $H^\mathrm{eff}$
only contains powers in $\sigma_x$ and $\sigma_y$ 
which are in total even, besides arbitrary powers in $\sigma_z$.
Because an even power in  $\sigma_x$ or $\sigma_y$ equals 
the identity and $\sigma_x\sigma_y=i\sigma_z$ we know 
\begin{equation}
\label{eq:hameff}
H^\mathrm{eff} = \sum_{\gamma\in\{0,z\}} \sigma_\gamma\otimes 
A_\gamma^\mathrm{eff},
\end{equation}
i.e., it only contains dephasing terms.
The order of the effective bath operators obviously read
\begin{subequations}
\label{eq:normalphaeff}
\begin{eqnarray}
||A_0^\mathrm{eff}||
&=& {\cal O}(\max(||A_0||,t ||A_x||^2,t ||A_y||^2))
\\
||A_z^\mathrm{eff}||
&=& 
{\cal O}(\max(||A_z||,t ||A_x||||A_y||)).
\end{eqnarray}
\end{subequations}
Note that the   $A_\gamma^\mathrm{eff}$ are complicated, non-linear functions
of $t $ due to the control sequence.

On this level, the idea to build another DD sequence based on
\eqref{eq:hameff} is very appealing because it is a pure dephasing 
Hamiltonian. The seemingly most efficient way is to use another UDD 
sequence on top of \eqref{eq:udd1}, but now
about the $x$- or the $y$-direction.
Unfortunately, this is not possible since $H^\mathrm{eff}$ is not independent
of $t $. Hence, if the UDD sequence requires the Hamiltonian
to be applied for a given duration this cannot be done by choosing some
interval $t $. The dependence of $H^\mathrm{eff}$ on 
$t $ is highly non-trival because it is non-linear and because
$[H^\mathrm{eff}(t ),H^\mathrm{eff}(t' )]\neq 0$ for 
$t  \neq t' $.

The situation is better if one is not aiming at arbitrary intervals but 
 at commensurate intervals $\{t_j\}$ which can be made from integer multiples 
of a basic interval $\Delta t$ like $t_j =n_j \Delta t$ with 
$n_j\in\mathbb{N}$.
Then one may choose $t =\Delta t$ and generate
$ \exp(-in_j t  H^\mathrm{eff})$ by applying the UDD
sequence $n_j$ times. But the intervals $\tilde \delta_{j+1}-
\tilde \delta_{j}$ are in general not commensurate, but see below.
Only its most basic versions, the spin-echo for $n=1$,
and the CPMG cycle for $n=2$ only use commensurate intervals.
This is already an interesting observation since the CPMG sequence 
$p_\mathrm{CPMG}$ suppresses  dephasing up to order $t^3$
as can be seen by regarding CPMG as the $n=2$ case of either 
UDD \cite{uhrig07} or CDD, see also below.

To assess the possible gain in concatenating CPMG with a UDD sequence
$p^m_\mathrm{UDD}$ of $m$ pulses and duration $t_1$
we estimate that the neglected term in \eqref{eq:efftimevol} is of the order
of $(\alpha t_1)^{m+1}$
This is the error committed on the UDD level of the sequence. On the
next level we consider 
\begin{equation}
p_\mathrm{CPMG}=p^m_\mathrm{UDD} X p^m_\mathrm{UDD}p^m_\mathrm{UDD} X
p^m_\mathrm{UDD}
\end{equation}
so that the total duration is $t=4t_1$. Then we have
$p_\mathrm{CPMG}=\mathbbm{1}+R$ where the deviation $R$
is estimated to be of the order of 
 $||R||= (\alpha^\mathrm{eff} t/8)^3$ with
$\alpha^\mathrm{eff}:=\max_\gamma(||A^\mathrm{eff}_\gamma||)$.
The validity of the factor $1/8$ becomes obvious in the 
generalization of the CPMG to arbitrary level of concatenation below.

So the total deviation $R^\mathrm{tot}_\mathrm{CPMG-UDD}$ will be approximately
\begin{equation}
\label{eq:rtot-cpmg}
||R^\mathrm{tot}_\mathrm{CPMG-UDD}|| \approx 
\max((\alpha^\mathrm{eff} t/8)^3,(\alpha t/4)^{m+1}).
\end{equation}
As expected one has to compare expressions involving different powers of
$t$. The use of $m>2$ is justified if $\alpha^\mathrm{eff}$ is much smaller
than $2\alpha$. In view of \eqref{eq:normalphaeff} this will be the case
if $\alpha$ and $\alpha^\mathrm{eff}$ are dominated by the $||A_\gamma||$
with $\gamma\in\{x,y,z\}$ and  if $||A_z||\ll||A_x||$ and $||A_z||\ll||A_y||$.
This observation shows that already a single concatenation step
on top of the optimized UDD sequences can be very useful.
Moreover, it is possible to pass to iterated CPMG sequences concatenated 
with basic UDD building blocks as in
$p_\mathrm{iCPMG}=(p^m_\mathrm{UDD} X p^m_\mathrm{UDD})^{2c}$ 
with $t_1=t/(2c)$, 
where $c$ is the number of iteration cycles.
Though the overvall order in the duration is not changed by the iteration
the influence of the unwanted couplings which are averaged to zero becomes
smaller and smaller on increasing $c$. This is so because the intervals between
pulses become smaller and smaller, see for instance Ref.\ \onlinecite{uhrig08}
for data for the bosonic dephasing model.

If the order $t^3$ of the CPMG is insufficient further levels of concatenations
can be added to reach higher orders. 
To see this it suffices to consider the simplemost
building block of concatenation \cite{khodj07}
\begin{equation}
\label{eq:simplerecurs}
p_{n+1}=p_{n}Xp_{n}X
\end{equation}
If the time evolution $p_n$ is governed by 
$H^{(n)}=\sum_{\gamma\in\{0,z\}} \sigma_\gamma\otimes A_\gamma^{(n)}$
the Magnus expansion \cite{magnu54} straightforwardly implies the recursion
\begin{subequations}
\label{eq:recursion}
\begin{eqnarray}
\label{eq:recurs-tau}
\tau_n &=& 2\tau_{n-1}\\
A_0^{(n+1)} &=& A_0^{(n)} + 
{\cal O}(\alpha_{n}\tau_{n}^2) \\
A_z^{(n+1)} &=& i(\tau_{n}/2)[A_0^{(n)}, A_z^{(n)}] + 
{\cal O}(\alpha_{n}\tau_{n}^2)
\label{eq:recurs-essential}
\end{eqnarray}
\end{subequations}
between consecutive levels of concatenation where $\tau_n$ is the total
duration of $p_n$ and $\alpha_{n}:=\max_{\gamma\in\{0,z\}}||A_\gamma^{(n)}||$.

Several remarks are in order: (i) The twofold concatenation 
according to \eqref{eq:simplerecurs} yields the CPMG
sequence: $p_2 = p_0 X p_0^2 X p_0$ because
$X^2=\mathbbm{1}$. (ii) If one starts with bath operators
$A_0$ and $A_z$ no terms in $x$- or $y$-direction are generated.
(iii) Note that it is sufficent to consider the leading orders of
the Magnus expansion  to derive \eqref{eq:recursion} because 
we are only interested in  the leading deviations.
(iv) The number $a_n$ of $X$ pulses grows roughly like $2^n$. More precisely,
it grows like $a_{n+1}= 2a_n+2(-1)^n$. Starting from $a_0=0$ this implies 
the exponential law $a_n=(2/3)(2^n-(-1)^n)$.

The essential point is that in $z$-direction each level of concatenation 
introduces an additional factor of $\tau_n$, see \eqref{eq:recurs-essential}. 
So each level of concatenation 
suppresses decoherence by an additonal order in the total duration of
the pulse sequence. The pure bath operator $A_0^{(n)}$ 
does not affect the qubit. This kind of property was exploited
in Refs.\ \onlinecite{khodj05,khodj07} to establish CDD. 
Hence, if we concatenate $n$ times according to \eqref{eq:simplerecurs}
we will achieve a sizable suppression of decoherence due to high 
powers in the total duration. 

More quantitatively, instead of \eqref{eq:rtot-cpmg} we have
\begin{equation}
\label{eq:rtot-concat}
||R^\mathrm{tot}_\mathrm{CUDD}|| \approx
\max\left((\alpha^\mathrm{eff} t 2^{\frac{-n(n+1)}{2}})^{n+1},
(\alpha t 2^{-n})^{m+1} \right),
\end{equation}
where $t$ is the total duration of the entire sequence.
The factor $2^{-n(n+1)/2}$ in the first term results from the product
of all the $\tau_n$ obeying the recursion \eqref{eq:recurs-tau}.
The factor $2^{-n}$ in the second term results from the number $2^n$
of UDD basic building blocks; each UDD interval has length $t 2^{-n}$.
Taking \eqref{eq:normalphaeff} into account we can 
conclude that $\alpha\leq\alpha^\mathrm{eff}$
for short enough times. Then we arrive at the
following conservative, but simple, estimate 
$||R^\mathrm{tot}_\mathrm{CUDD}|| \approx (\alpha t 2^{-m})^{m+1})$ 
for $n=m$. Hence we realize that the combination of concatenation and interval 
optimization makes it indeed possible to reduce any type of coupling
to high order.

In assessing the gain of the advocated CUDD 
over the known CDD approach, we have to 
consider the number of necessary pulses. In order to make the $t^m$ term 
vanish the CDD approach requires $4^m$ pulses \cite{khodj05,khodj07}.
The CUDD (concatenated UDD) sequence on concatenation level $n$ made from
UDD sequences with $m$ pulses requires 
$a_n\approx (2/3)(2^n-(-1)^n)$ pulses of type $X$ and $m2^n$ pulses
of type $Z$. To suppress all $t^m$ terms we choose $n=m$ and the total
number of pulses is about $(m+2/3)2^m$. Thus we have achieved a substantial
reduction by passing from $4^m$  to $(m+2/3)2^m$ if we replace CDD by CUDD.
So one factor $2^m$ is reduced to the linear factor $m+2/3$ due to the UDD
building blocks. Note that this pays already for small $m\ge 1$.
This is the main general result of the present work.

Unfortunately, the suppression of the dephasing
component along $\sigma_z$ could not be reduced in the same fashion
from exponential to linear by employing UDD 
due to commensurability problems, see above. If one were able to use 
a UDD scheme also for the second step the number of pulses could be
brought down to the order $m^2$. In practice, it can be promising
to check out a work-around. The function $d(x)=\sin^2(\pi x/2)$ 
on the right hand side of \eqref{eq:udd0} ($x=j/(n+1)$) 
has a simple shape,
see Fig.\ 1. It is odd about $(1/2,1/2)$, it vanishes at $x=0$, and it has
vanishing slope  at $x=0$ and $x=1$. These features are reproduced by a simple
third order polynomial $d_\mathrm{approx}=-2x^3+3x^2$ which approximates
$d(x)$ very well as shown in Fig.\ 1. 
The maximum difference is only about $0.01$ so that it is to be assumed
that the use of $d_\mathrm{approx}(x)$ instead of $d(x)$ is hardly
noticeable in practice.
\begin{figure}[ht]
    \begin{center}
     \includegraphics[width=0.98\columnwidth,clip]{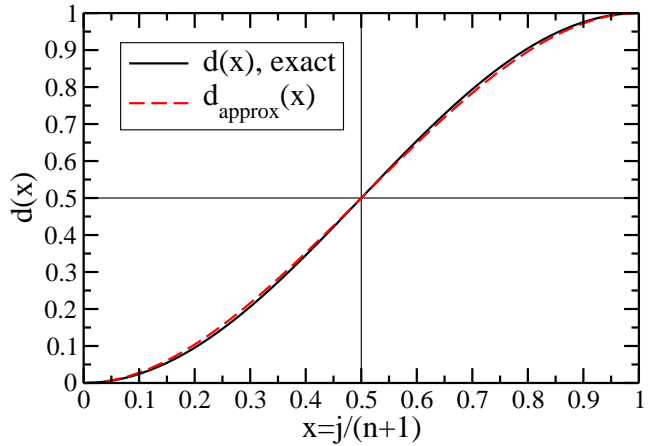}
    \end{center}
    \caption{(color online) Right hand side $d(x)=\sin^2(\pi x/2)$  of Eq.\ 
      \eqref{eq:udd0} compared to $d_\mathrm{approx}=-2x^3+3x^2$.}
\end{figure}

The use of $d_\mathrm{approx}(x)$ recovers commensurability. Because
 $x=j/(n+1)$
and the third power in $d_\mathrm{approx}(x)$ is the highest power occurring,
all instants $\tilde \delta_j=d_\mathrm{approx}(j/(n+1))$ are integer
multiples of $t/(n+1)^3$\footnote{If $n+1=2l$ with $l$ odd, all instants
$\tilde \delta_j$ are even integer multiples of $4/(n+1)^3$.}.
This means that we have to realize $(n+1)^3$ intervals of UDD sequences
with $n$ $Z$-pulses and at the $n$ instants $\tilde \delta_j$
another $n$ $X$-pulses are needed. Hence this type of UDD sequence on top
of a UDD sequence (approximate UDD$^2$) requires $n(n+1)^3+n$ pulses, 
hence only a polynomial law as opposed to the CUDD which still
requires an exponential number of pulses. Due to the high power
of the polynomial (degree 4), the approximate UDD$^2$ pays only for
larger values of $n$ if $(n+1)^3\lessapprox 2^n$, i.e., for $n\gtrapprox 10$.
Further exploration of the approximate UDD$^2$ by numerical means is called
for because of the approximation involved which precludes a rigorous
analytic assessment.

In summary, we have shown that UDD can be used as a powerful building block
in more complex pulse sequences. Thereby the restriction of being applicable
only to the suppression of either pure dephasing or pure relaxation
is abolished. Three schemes are proposed. The first is suited for
 anisotropic situations where
spin flips are much less likely than dephasing. The UDD can be inserted in
a CPMG cycle. The CPMG can also be iterated if needed. This scheme
yields a suppression in $t^3$ only due to the properties of the CPMG. But
in view of the robust properties of the CPMG cycle this may be fully
sufficient in many circumstances.

The second, most general and flexible, 
scheme consists in using the UDD sequence as 
starting sequence of concatenation to achieve a concatenated 
UDD (CUDD). This scheme makes it possible to suppress 
decoherence to arbitrary order $t^{m+1}$ in the total duration $t$
if a UDD sequence of $m$ pulses is concatenated $m$ times. The
necessary number of pulse scales like $m2^m$ which is
still exponential but considerably less demanding than $4^m$ as
in CDD.

Third, an approximate scheme of two concatenated UDD sequences is
proposed. The appoximation is needed on the second level to retrieve a
certain degree of commensurability. The required number of pulses scales
like $m^4$ so that it becomes more advantageous than CUDD for
a sufficiently large number of pulses ($m\gtrapprox 10$).

The proposed schemes are expected to be experimentally useful
for high-precision NMR where narrow line-widths are a prerequisite
or for long-time data storage in realizations of quantum information
devices.

\acknowledgments

I like to thank S.E.\ Barrett and O.P.\ Sushkov for helpful discussions.
The financial support by  the Heinrich-Hertz Stiftung NRW and by the Gordon 
Godfrey Fund is gratefully acknowledged.


\begin{thebibliography}{25}
\expandafter\ifx\csname natexlab\endcsname\relax\def\natexlab#1{#1}\fi
\expandafter\ifx\csname bibnamefont\endcsname\relax
  \def\bibnamefont#1{#1}\fi
\expandafter\ifx\csname bibfnamefont\endcsname\relax
  \def\bibfnamefont#1{#1}\fi
\expandafter\ifx\csname citenamefont\endcsname\relax
  \def\citenamefont#1{#1}\fi
\expandafter\ifx\csname url\endcsname\relax
  \def\url#1{\texttt{#1}}\fi
\expandafter\ifx\csname urlprefix\endcsname\relax\def\urlprefix{URL }\fi
\providecommand{\bibinfo}[2]{#2}
\providecommand{\eprint}[2][]{\url{#2}}

\bibitem[{\citenamefont{Haeberlen}(1976)}]{haebe76}
\bibinfo{author}{\bibfnamefont{U.}~\bibnamefont{Haeberlen}},
  \emph{\bibinfo{title}{High Resolution NMR in Solids: Selective Averaging}}
  (\bibinfo{publisher}{Academic Press}, \bibinfo{address}{New York},
  \bibinfo{year}{1976}).

\bibitem[{\citenamefont{Zoller et~al.}(2005)\citenamefont{Zoller, Beth, Binosi,
  Blatt, Briegel, Bruss, Calarco, Cirac, Deutsch, Eisert et~al.}}]{zolle05}
\bibinfo{author}{\bibfnamefont{P.}~\bibnamefont{Zoller}},
  \bibinfo{author}{\bibfnamefont{T.}~\bibnamefont{Beth}},
  \bibinfo{author}{\bibfnamefont{D.}~\bibnamefont{Binosi}},
  \bibinfo{author}{\bibfnamefont{R.}~\bibnamefont{Blatt}},
  \bibinfo{author}{\bibfnamefont{H.}~\bibnamefont{Briegel}},
  \bibinfo{author}{\bibfnamefont{D.}~\bibnamefont{Bruss}},
  \bibinfo{author}{\bibfnamefont{T.}~\bibnamefont{Calarco}},
  \bibinfo{author}{\bibfnamefont{J.~I.} \bibnamefont{Cirac}},
  \bibinfo{author}{\bibfnamefont{D.}~\bibnamefont{Deutsch}},
  \bibinfo{author}{\bibfnamefont{J.}~\bibnamefont{Eisert}},
  \bibnamefont{et~al.}, \bibinfo{journal}{Eur. Phys. J. D}
  \textbf{\bibinfo{volume}{36}}, \bibinfo{pages}{203} (\bibinfo{year}{2005}).

\bibitem[{\citenamefont{Steane}(1998)}]{stean98}
\bibinfo{author}{\bibfnamefont{A.}~\bibnamefont{Steane}},
  \bibinfo{journal}{Rep. Prog. Phys.} \textbf{\bibinfo{volume}{61}},
  \bibinfo{pages}{117} (\bibinfo{year}{1998}).

\bibitem[{\citenamefont{Viola and Lloyd}(1998)}]{viola98}
\bibinfo{author}{\bibfnamefont{L.}~\bibnamefont{Viola}} \bibnamefont{and}
  \bibinfo{author}{\bibfnamefont{S.}~\bibnamefont{Lloyd}},
  \bibinfo{journal}{Phys. Rev. A} \textbf{\bibinfo{volume}{58}},
  \bibinfo{pages}{2733} (\bibinfo{year}{1998}).

\bibitem[{\citenamefont{Ban}(1998)}]{ban98}
\bibinfo{author}{\bibfnamefont{M.}~\bibnamefont{Ban}}, \bibinfo{journal}{J.
  Mod. Opt.} \textbf{\bibinfo{volume}{45}}, \bibinfo{pages}{2315}
  (\bibinfo{year}{1998}).

\bibitem[{\citenamefont{Facchi et~al.}(2005)\citenamefont{Facchi, Tasaki,
  Pascazio, Nakazato, Tokuse, and Lidar}}]{facch05}
\bibinfo{author}{\bibfnamefont{P.}~\bibnamefont{Facchi}},
  \bibinfo{author}{\bibfnamefont{S.}~\bibnamefont{Tasaki}},
  \bibinfo{author}{\bibfnamefont{S.}~\bibnamefont{Pascazio}},
  \bibinfo{author}{\bibfnamefont{H.}~\bibnamefont{Nakazato}},
  \bibinfo{author}{\bibfnamefont{A.}~\bibnamefont{Tokuse}}, \bibnamefont{and}
  \bibinfo{author}{\bibfnamefont{D.~A.} \bibnamefont{Lidar}},
  \bibinfo{journal}{Phys. Rev. A} \textbf{\bibinfo{volume}{71}},
  \bibinfo{pages}{022302} (\bibinfo{year}{2005}).

\bibitem[{\citenamefont{Viola et~al.}(1999)\citenamefont{Viola, Knill, and
  Lloyd}}]{viola99a}
\bibinfo{author}{\bibfnamefont{L.}~\bibnamefont{Viola}},
  \bibinfo{author}{\bibfnamefont{E.}~\bibnamefont{Knill}}, \bibnamefont{and}
  \bibinfo{author}{\bibfnamefont{S.}~\bibnamefont{Lloyd}},
  \bibinfo{journal}{Phys. Rev. Lett.} \textbf{\bibinfo{volume}{82}},
  \bibinfo{pages}{2417} (\bibinfo{year}{1999}).

\bibitem[{\citenamefont{Khodjasteh and Lidar}(2005)}]{khodj05}
\bibinfo{author}{\bibfnamefont{K.}~\bibnamefont{Khodjasteh}} \bibnamefont{and}
  \bibinfo{author}{\bibfnamefont{D.~A.} \bibnamefont{Lidar}},
  \bibinfo{journal}{Phys. Rev. Lett.} \textbf{\bibinfo{volume}{95}},
  \bibinfo{pages}{180501} (\bibinfo{year}{2005}).

\bibitem[{\citenamefont{Uhrig}(2007)}]{uhrig07}
\bibinfo{author}{\bibfnamefont{G.~S.} \bibnamefont{Uhrig}},
  \bibinfo{journal}{Phys. Rev. Lett.} \textbf{\bibinfo{volume}{98}},
  \bibinfo{pages}{100504} (\bibinfo{year}{2007}).

\bibitem[{\citenamefont{Hahn}(1950)}]{hahn50}
\bibinfo{author}{\bibfnamefont{E.~L.} \bibnamefont{Hahn}},
  \bibinfo{journal}{Phys. Rev.} \textbf{\bibinfo{volume}{80}},
  \bibinfo{pages}{580} (\bibinfo{year}{1950}).

\bibitem[{\citenamefont{Carr and Purcell}(1954)}]{carr54}
\bibinfo{author}{\bibfnamefont{H.~Y.} \bibnamefont{Carr}} \bibnamefont{and}
  \bibinfo{author}{\bibfnamefont{E.~M.} \bibnamefont{Purcell}},
  \bibinfo{journal}{Phys. Rev.} \textbf{\bibinfo{volume}{94}},
  \bibinfo{pages}{630} (\bibinfo{year}{1954}).

\bibitem[{\citenamefont{Meiboom and Gill}(1958)}]{meibo58}
\bibinfo{author}{\bibfnamefont{S.}~\bibnamefont{Meiboom}} \bibnamefont{and}
  \bibinfo{author}{\bibfnamefont{D.}~\bibnamefont{Gill}},
  \bibinfo{journal}{Rev. Sci. Inst.} \textbf{\bibinfo{volume}{29}},
  \bibinfo{pages}{688} (\bibinfo{year}{1958}).

\bibitem[{\citenamefont{Witzel and DasSarma}(2007)}]{witze07a}
\bibinfo{author}{\bibfnamefont{W.~M.} \bibnamefont{Witzel}} \bibnamefont{and}
  \bibinfo{author}{\bibfnamefont{S.}~\bibnamefont{DasSarma}},
  \bibinfo{journal}{Phys. Rev. Lett.} \textbf{\bibinfo{volume}{98}},
  \bibinfo{pages}{077601} (\bibinfo{year}{2007}).

\bibitem[{\citenamefont{Yao et~al.}(2007)\citenamefont{Yao, Liu, and
  Sham}}]{yao07}
\bibinfo{author}{\bibfnamefont{W.}~\bibnamefont{Yao}},
  \bibinfo{author}{\bibfnamefont{R.~B.} \bibnamefont{Liu}}, \bibnamefont{and}
  \bibinfo{author}{\bibfnamefont{L.~J.} \bibnamefont{Sham}},
  \bibinfo{journal}{Phys. Rev. Lett.} \textbf{\bibinfo{volume}{98}},
  \bibinfo{pages}{077602} (\bibinfo{year}{2007}).

\bibitem[{\citenamefont{Uhrig}(2008)}]{uhrig08}
\bibinfo{author}{\bibfnamefont{G.~S.} \bibnamefont{Uhrig}},
  \bibinfo{journal}{New J. Phys.} \textbf{\bibinfo{volume}{10}},
  \bibinfo{pages}{083024} (\bibinfo{year}{2008}).

\bibitem[{\citenamefont{Cywi\'{n}ski et~al.}(2008)\citenamefont{Cywi\'{n}ski,
  Lutchyn, Nave, and DasSarma}}]{cywin08}
\bibinfo{author}{\bibfnamefont{L.}~\bibnamefont{Cywi\'{n}ski}},
  \bibinfo{author}{\bibfnamefont{R.~M.} \bibnamefont{Lutchyn}},
  \bibinfo{author}{\bibfnamefont{C.~P.} \bibnamefont{Nave}}, \bibnamefont{and}
  \bibinfo{author}{\bibfnamefont{S.}~\bibnamefont{DasSarma}},
  \bibinfo{journal}{Phys. Rev. B} \textbf{\bibinfo{volume}{77}},
  \bibinfo{pages}{174509} (\bibinfo{year}{2008}).

\bibitem[{\citenamefont{Khodjasteh and Lidar}(2007)}]{khodj07}
\bibinfo{author}{\bibfnamefont{K.}~\bibnamefont{Khodjasteh}} \bibnamefont{and}
  \bibinfo{author}{\bibfnamefont{D.~A.} \bibnamefont{Lidar}},
  \bibinfo{journal}{Phys. Rev. A} \textbf{\bibinfo{volume}{75}},
  \bibinfo{pages}{062310} (\bibinfo{year}{2007}).

\bibitem[{\citenamefont{Dhar et~al.}(2006)\citenamefont{Dhar, Grover, and
  Roy}}]{dhar06}
\bibinfo{author}{\bibfnamefont{D.}~\bibnamefont{Dhar}},
  \bibinfo{author}{\bibfnamefont{L.~K.} \bibnamefont{Grover}},
  \bibnamefont{and} \bibinfo{author}{\bibfnamefont{S.~M.} \bibnamefont{Roy}},
  \bibinfo{journal}{Phys. Rev. Lett.} \textbf{\bibinfo{volume}{96}},
  \bibinfo{pages}{100405} (\bibinfo{year}{2006}).

\bibitem[{\citenamefont{Lee et~al.}(2008)\citenamefont{Lee, Witzel, and
  DasSarma}}]{lee08a}
\bibinfo{author}{\bibfnamefont{B.}~\bibnamefont{Lee}},
  \bibinfo{author}{\bibfnamefont{W.~M.} \bibnamefont{Witzel}},
  \bibnamefont{and} \bibinfo{author}{\bibfnamefont{S.}~\bibnamefont{DasSarma}},
  \bibinfo{journal}{Phys. Rev. Lett.} \textbf{\bibinfo{volume}{100}},
  \bibinfo{pages}{160505} (\bibinfo{year}{2008}).

\bibitem[{\citenamefont{Yang and Liu}(2008)}]{yang08}
\bibinfo{author}{\bibfnamefont{W.}~\bibnamefont{Yang}} \bibnamefont{and}
  \bibinfo{author}{\bibfnamefont{R.-B.} \bibnamefont{Liu}},
  \bibinfo{journal}{Phys. Rev. Lett.} \textbf{\bibinfo{volume}{101}},
  \bibinfo{pages}{180403} (\bibinfo{year}{2008}).

\bibitem[{\citenamefont{Li et~al.}(2007)\citenamefont{Li, Dementyev, Dong,
  Ramos, and Barrett}}]{li07}
\bibinfo{author}{\bibfnamefont{D.}~\bibnamefont{Li}},
  \bibinfo{author}{\bibfnamefont{A.~E.} \bibnamefont{Dementyev}},
  \bibinfo{author}{\bibfnamefont{Y.}~\bibnamefont{Dong}},
  \bibinfo{author}{\bibfnamefont{R.~G.} \bibnamefont{Ramos}}, \bibnamefont{and}
  \bibinfo{author}{\bibfnamefont{S.~E.} \bibnamefont{Barrett}},
  \bibinfo{journal}{Phys. Rev. Lett.} \textbf{\bibinfo{volume}{98}},
  \bibinfo{pages}{190401} (\bibinfo{year}{2007}).

\bibitem[{\citenamefont{Li et~al.}(2008)\citenamefont{Li, Dong, Ramos, Murray,
  MacLean, Dementyev, and Barrett}}]{li08a}
\bibinfo{author}{\bibfnamefont{D.}~\bibnamefont{Li}},
  \bibinfo{author}{\bibfnamefont{Y.}~\bibnamefont{Dong}},
  \bibinfo{author}{\bibfnamefont{R.~G.} \bibnamefont{Ramos}},
  \bibinfo{author}{\bibfnamefont{J.~D.} \bibnamefont{Murray}},
  \bibinfo{author}{\bibfnamefont{K.}~\bibnamefont{MacLean}},
  \bibinfo{author}{\bibfnamefont{A.~E.} \bibnamefont{Dementyev}},
  \bibnamefont{and} \bibinfo{author}{\bibfnamefont{S.~E.}
  \bibnamefont{Barrett}}, \bibinfo{journal}{Phys. Rev. B}
  \textbf{\bibinfo{volume}{77}}, \bibinfo{pages}{214306}
  (\bibinfo{year}{2008}).

\bibitem[{\citenamefont{Pasini et~al.}(2008)\citenamefont{Pasini, Fischer,
  Karbach, and Uhrig}}]{pasin08a}
\bibinfo{author}{\bibfnamefont{S.}~\bibnamefont{Pasini}},
  \bibinfo{author}{\bibfnamefont{T.}~\bibnamefont{Fischer}},
  \bibinfo{author}{\bibfnamefont{P.}~\bibnamefont{Karbach}}, \bibnamefont{and}
  \bibinfo{author}{\bibfnamefont{G.~S.} \bibnamefont{Uhrig}},
  \bibinfo{journal}{Phys. Rev. A} \textbf{\bibinfo{volume}{77}},
  \bibinfo{pages}{032315} (\bibinfo{year}{2008}).

\bibitem[{\citenamefont{Karbach et~al.}(2008)\citenamefont{Karbach, Pasini, and
  Uhrig}}]{karba08}
\bibinfo{author}{\bibfnamefont{P.}~\bibnamefont{Karbach}},
  \bibinfo{author}{\bibfnamefont{S.}~\bibnamefont{Pasini}}, \bibnamefont{and}
  \bibinfo{author}{\bibfnamefont{G.~S.} \bibnamefont{Uhrig}},
  \bibinfo{journal}{Phys. Rev. A} \textbf{\bibinfo{volume}{78}},
  \bibinfo{pages}{022315} (\bibinfo{year}{2008}).

\bibitem[{\citenamefont{Magnus}(1954)}]{magnu54}
\bibinfo{author}{\bibfnamefont{W.}~\bibnamefont{Magnus}},
  \bibinfo{journal}{Comm. Pure Appl. Math.} \textbf{\bibinfo{volume}{7}},
  \bibinfo{pages}{649} (\bibinfo{year}{1954}).

\end{thebibliography}

\end{document}